\documentclass[11pt]{cernrep}
\usepackage{graphicx,epsfig,epsf,psfig,amssymb,
rotating,colordvi,helvet,color,subfigure,axodraw}
\def\B0bar{\overline{B^0}}

\begin{document}

 \title{{\normalsize\rm{DFTT 1/2004\hskip11.85cm{SHEP-03-39}}}\\
ELECTROWEAK RADIATIVE CORRECTIONS TO HADRONIC PRECISION OBSERVABLES
AT TEV ENERGIES}
\author{E. Maina$^1$, S. Moretti$^2$, M.R. Nolten$^2$ and D.A. Ross$^2$}
\institute{$^1$INFN and Universit\`a di Torino, Italy; $^2$Southampton University, UK}
\maketitle
\begin{abstract}
We illustrate the impact of full one-loop weak corrections onto
$b$-jet-, prompt-photon- and $Z$-production at Tevatron and
Large Hadron Collider (LHC).
\end{abstract}

\section{WEAK CORRECTIONS AT TEV SCALES}

At TeV energy scales, next-to-leading order (NLO) 
Electro-Weak (EW) effects produce leading
corrections of the type $\alpha_{\rm{EW}}\log^2({\hat{s}}/M_W^2)$, where 
$\alpha_{\mathrm{EW}}\equiv \alpha_{\mathrm{EM}}\sin^2\theta_W$,
with $\alpha_{\mathrm{EM}}$ the Electro-Magnetic coupling
constant and $\theta_W$ the Weinberg angle. In fact,
for large enough $\hat s$ values, the centre-of-mass (CM)
energy at parton level, such EW effects may be competitive not
only with next-to-NLO (NNLO) (as $ \alpha_{\rm{EW}}\approx 
\alpha_{\rm{S}}^2$) but also with NLO QCD corrections (e.g., for
$\sqrt{\hat{s}}=3$ TeV, $\log^2({\hat{s}}/M_W^2)\approx10$).

These `double logs' are 
due to a lack of cancellation between virtual and real $W$-emission in
higher order contributions. This is in turn a consequence of the 
violation of the Bloch-Nordsieck theorem in non-Abelian theories
\cite{Ciafaloni:2000df}.
The problem is in principle present also in QCD. In practice, however, 
it has no observable consequences, because of the final averaging of the 
colour degrees of freedom of partons, forced by their confinement
into colourless hadrons. This does not occur in the EW case,
where the initial state has a non-Abelian charge,
as in an initial quark doublet in proton-(anti)proton scattering. 
Besides, these
logarithmic corrections are finite (unlike in
QCD), since $M_W$ provides a physical
cut-off for $W$-emission. Hence, for typical experimental
resolutions, softly and collinearly emitted weak bosons need not be included
in the production cross section and one can restrict oneself to the 
calculation
of weak effects originating from virtual corrections only. 
By doing so, similar
logarithmic effects, $\sim\alpha_{\rm{EW}}\log^2({\hat{s}}/M_Z^2)$, 
are generated also by $Z$-boson corrections.
Finally, all these purely weak contributions can  be
isolated in a gauge-invariant manner from EM effects which therefore may not
be included in the calculation. In fact, we have neglected the latter here. 

In view of all this,  it becomes of crucial importance to assess
the quantitative relevance of such weak corrections
affecting, in particular, key QCD processes at present and future 
hadron colliders. We show here results for the case of $b$-jet-,
prompt-photon and $Z$-production at Tevatron and LHC.

\section{CORRECTIONS TO $\boldmath{b}$-JET-PRODUCTION}

\begin{figure}[!ht]
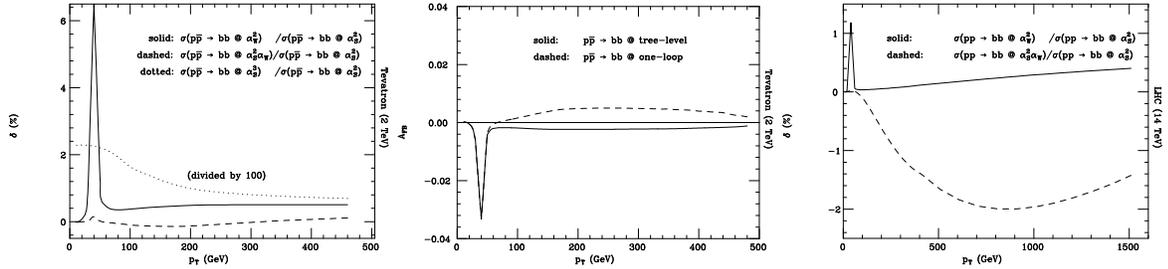

{\epsfig{file=ratiobb_Tev.ps,width=3.5cm,angle=90}}
{\epsfig{file=sigmabb_AFB_Tev.ps,width=3.5cm,angle=90}}
{\epsfig{file=ratiobb_LHC.ps,width=3.5cm,angle=90}}
\caption{The corrections (NLO-LO)/LO due to the 
$\alpha_{\rm{EW}}^2$, $\alpha_{\rm{S}}^2\alpha_{\rm{EW}}$ 
and $\alpha_{\rm{S}}^3$ terms
relative to the $\alpha_{\rm{S}}^2$ ones
vs. the transverse momentum
of the $b$-jet for $p\bar p\to b\bar b(g)$ and  $p p\to b\bar b(g)$ production at
Tevatron and LHC, left and right frame, respectively. 
(For LHC,
we do not show the corrections due to $\alpha_{\rm{S}}^3$ terms as results are
perturbatively unreliable.) In the middle frame, the forward-backward  
asymmetry vs. the transverse momentum
of the $b$-jet for $p\bar p\to b\bar b(g)$ events at
Tevatron, as obtained at tree-level ${\cal O}(\alpha_{\rm{EW}}^2)$ 
and one-loop ${\cal O}(\alpha_{\rm{S}}^2\alpha_{\rm{EW}})$ orders.}
\label{fig:b}
\vspace*{-0.5truecm}
\end{figure}

In Fig.~\ref{fig:b} (left and right panels) we show the effects of the full  
${\cal O}(\alpha_{\rm S}^2\alpha_{\rm{EW}})$ contributions to
the $p\bar p\to b\bar b(g)$ and  $pp\to b\bar b(g)$ cross sections
at Tevatron and LHC, respectively. (For details of the calculation,
see Ref.~\cite{Maina:2003is}.) Results are shown for the total inclusive
$b$-jet production rate as a function of the jet transverse
energy. (Tree-level EW and NLO QCD effects are also given for comparison.)
At Tevatron, ${\cal O}(\alpha_{\rm S}^2\alpha_{\rm{EW}})$ terms
are typically negligible in the inclusive cross section, as the partonic
energy available is too small for the mentioned logarithms to be 
effective. At LHC, the contribution due to such terms grows
accordingly to the collider energy, reaching the --2\% level at 
transverse momenta of $\approx800$ GeV. 

Next, we study the forward-backward asymmetry at Tevatron, defined as
follows:
\begin{equation}\label{AFB}
A^b_{\rm{FB}}=
\frac{\sigma_+[p\bar p\to b\bar b(g)]-\sigma_-[p\bar p\to b\bar b(g)]}
     {\sigma_+[p\bar p\to b\bar b(g)]+\sigma_-[p\bar p\to b\bar b(g)]},
\end{equation}
where the subscript $+(-)$ iden\-ti\-fies events in which the $b$-jet
is produced with polar angle larger(smal\-ler) than 90 degrees respect to 
one of the two beam directions (hereafter, we use
the proton beam as positive $z$-axis).
The polar angle is defined in the CM
frame of the hard partonic scattering. In the center plot of 
Fig.~\ref{fig:b}, the solid curve
represents the sum of the tree-level contributions, that is, 
those of order $\alpha_{\rm{S}}^2$ and  
$\alpha_{\rm{EW}}^2$, whereas the dashed one also includes the
higher-order ones 
$\alpha_{\rm{S}}^2\alpha_{\rm{EW}}$. The effects of the one-loop weak corrections on 
this observable are rather large, indeed comparable to the effects
through order $\alpha_{\rm{S}}^3$ \cite{Kuhn:1998kw}. 
In absolute terms, the asymmetry is of order $-4\%$ 
at the $W$, $Z$ resonance (i.e., for $p_T\approx M_W/2, M_Z/2$)
 and
fractions of percent elsewhere, hence it should be measurable
after the end of Run 2. We expect even larger effects at LHC, however,
some care is here necessary in order
to define the observable, which depends on
the configuration and efficiency of the experimental apparata 
(so we do not present the corresponding plot in this instance). 
The $\alpha_{\rm{S}}^3$ results presented here are from                                
             Ref.~\cite{Frixione:1996nh}.

\section{CORRECTIONS TO $\boldmath{\gamma}$- AND $\boldmath{Z}$-PRODUCTION}

The neutral-current processes ($V=\gamma,Z$)
\begin{equation}\label{procs_neutral}
q\bar q \to g V\quad{\rm{and}}\quad q(\bar q) g\to q(\bar q) V
\end{equation}
are two of the cleanest probes of the partonic content of (anti)protons,
in particular of antiquark and gluon
densities. In order to measure the latter it is necessary to study
the vector boson $p_T$ spectrum. That is, to
compute $V$ production in association with a jet 
(originated by either a quark or a gluon).
We briefly report here on the full one-loop results for processes
(\ref{procs_neutral}) obtained through
${\cal O}(\alpha_{\rm{S}}\alpha_{\rm{EW}}^2)$. (For technical
details of the calculation, see Ref.~\cite{Maina:2002wz}.)

\begin{figure}[!h]
\begin{minipage}{\textwidth}
\includegraphics[width=.5\linewidth]{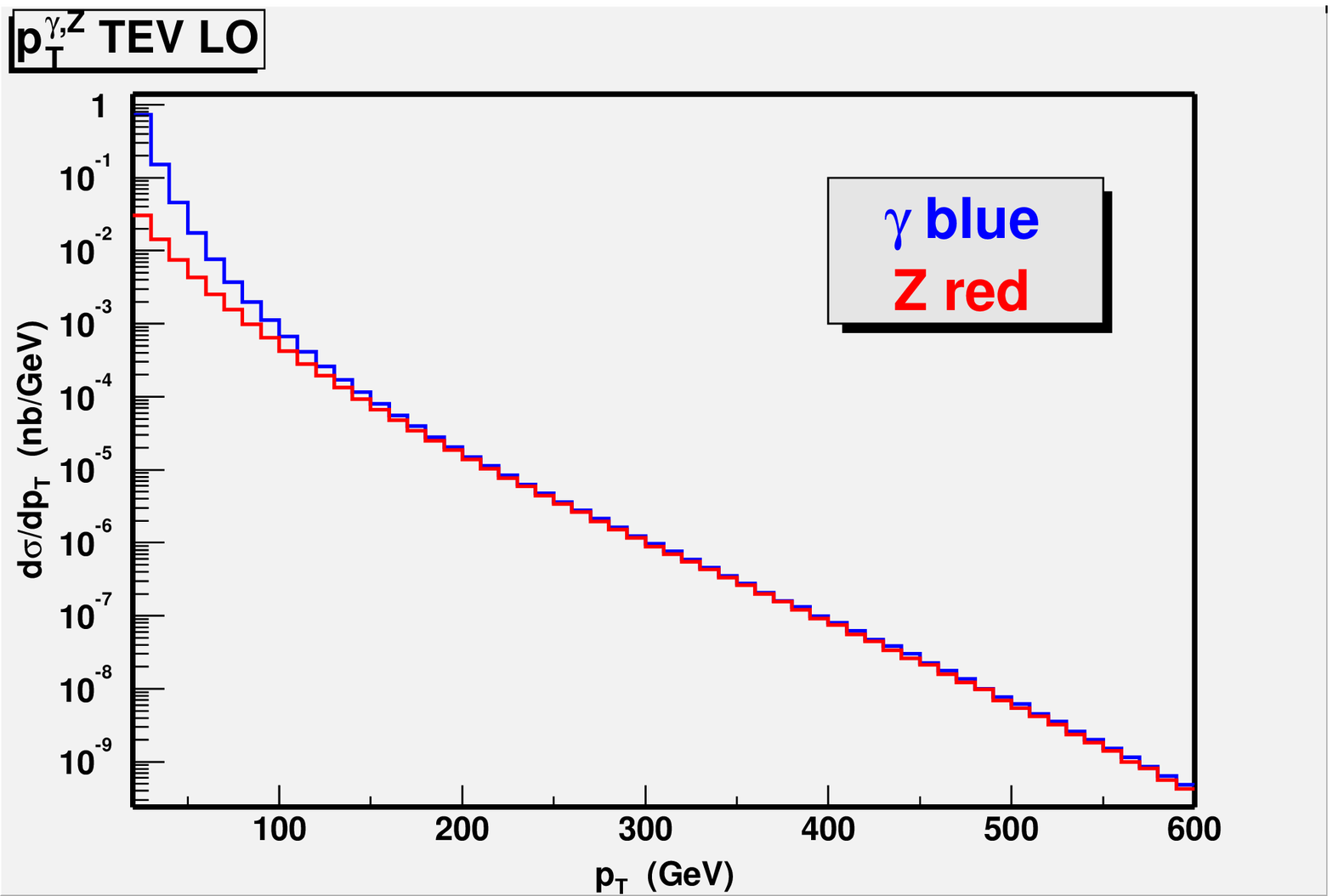}
\includegraphics[width=.5\linewidth]{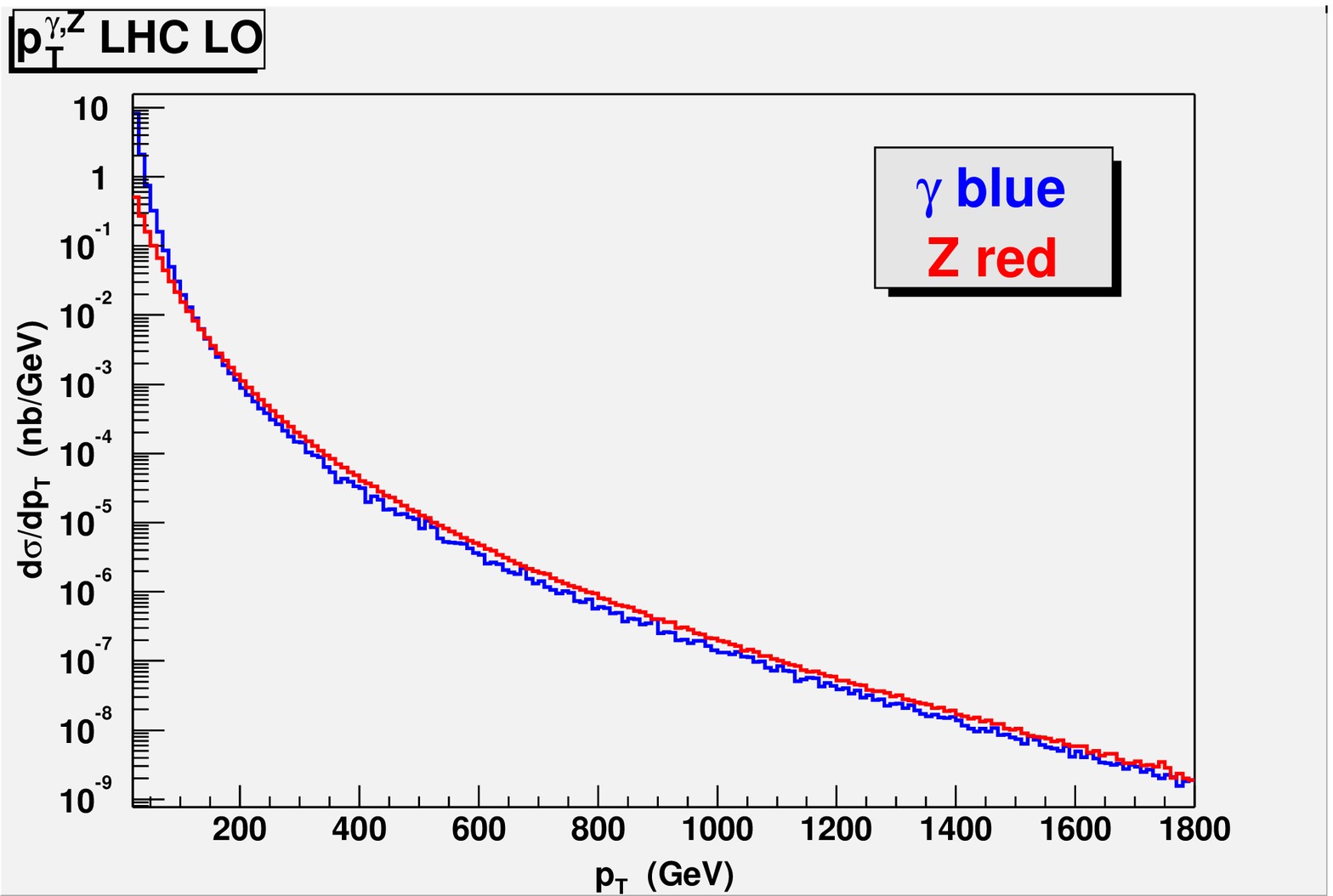}
\end{minipage}
\begin{minipage}{\textwidth}
\includegraphics[width=.5\linewidth]{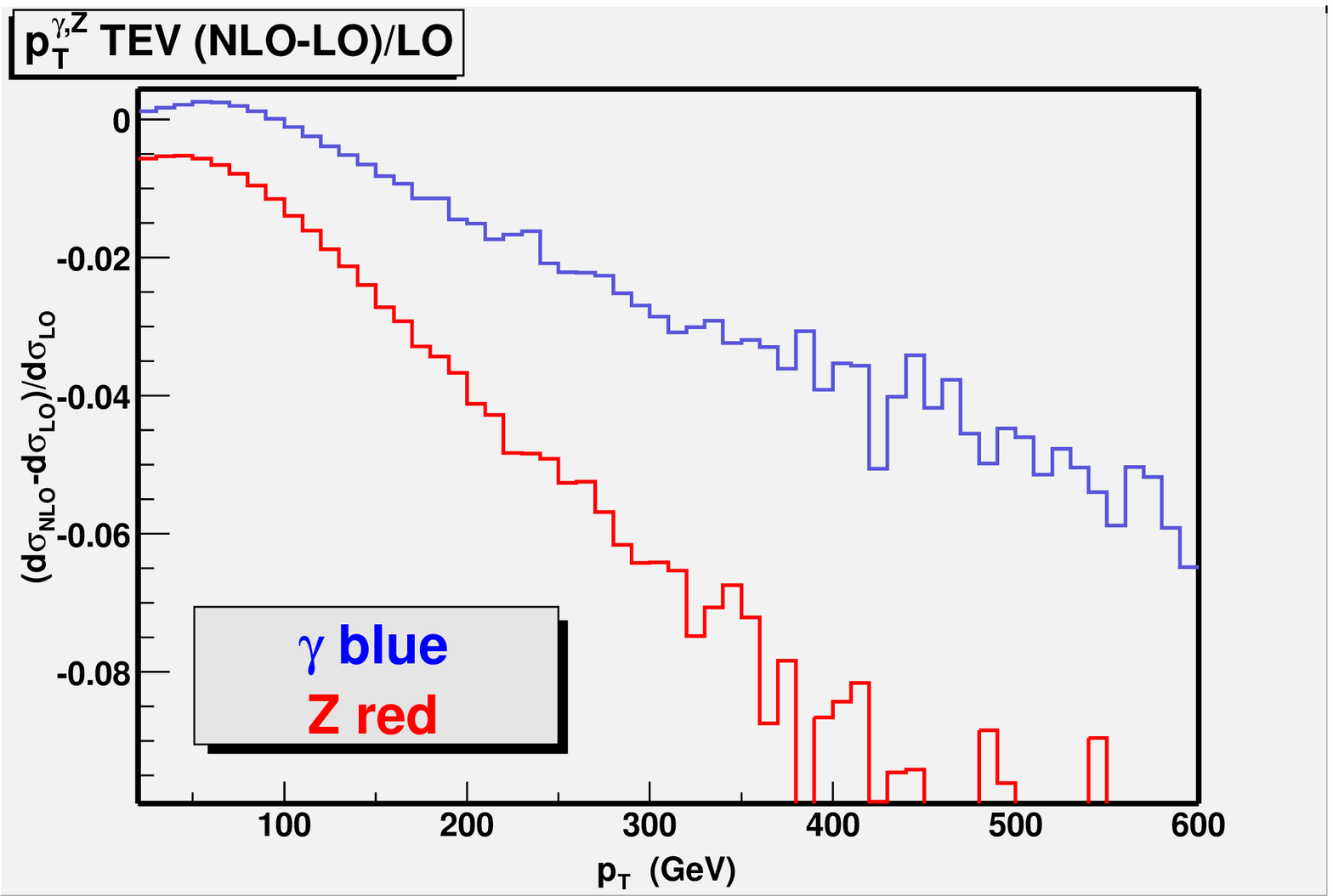}
\includegraphics[width=.5\linewidth]{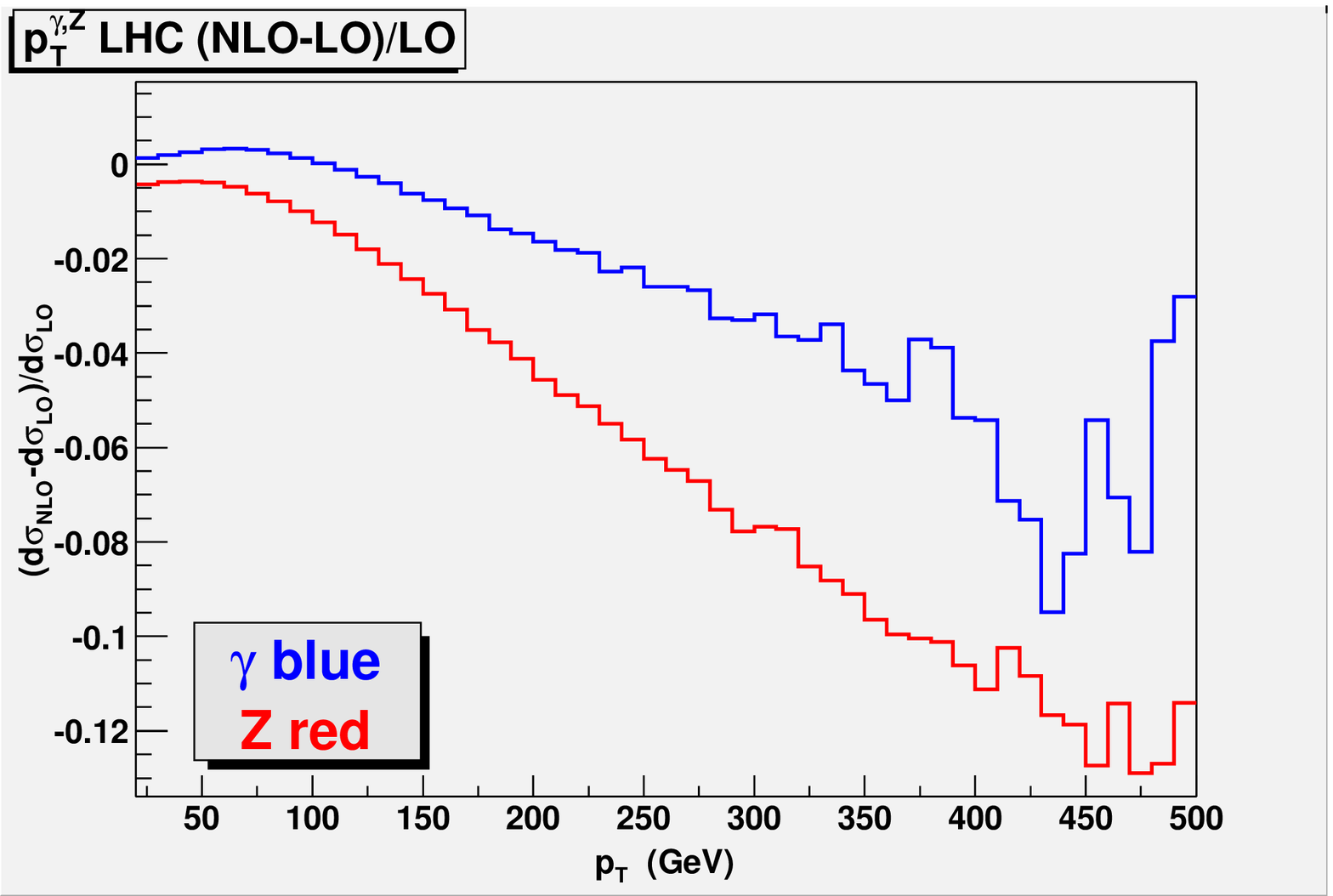}
\end{minipage}
\caption{The LO results through 
${\cal O}(\alpha_{\rm{S}}\alpha_{\rm{EW}})$
for the $\gamma$- and $Z$-production cross sections 
at Tevatron
and LHC, as a function of the transverse
momentum (top) as well as the size of the NLO 
corrections  through ${\cal O}(\alpha_{\rm{S}}\alpha_{\rm{EW}}^2)$
relatively to the former.} 
\label{fig:V}
\vspace*{-0.5truecm}
\end{figure}

Fig.~\ref{fig:V} shows the effects of the 
${\cal O}(\alpha_{\rm{S}}\alpha_{\rm{EW}}^2)$
corrections relatively to the ${\cal O}(\alpha_{\rm{S}}\alpha_{\rm{EW}})$
Born results ($\alpha_{\rm{EM}}$ replaces $\alpha_{\rm{EW}}$ for photons),
as well as the absolute magnitude of the latter, as a function
of the transverse momentum. The corrections are found to be rather
large, both at Tevatron and LHC, particularly
for $Z$-production. In case of the latter,
such effects are of order --10\% at Tevatron 
and --15\% at LHC for $p_T\approx 500$ GeV. In general, above 
$p_T\approx100$ GeV,
they tend to (negatively) increase, more or less linearly, with $p_T$.
Such effects are indeed observable at both Tevatron and LHC. 
For example, at FNAL, for $Z$-production and decay into electrons and muons
with BR$(Z\rightarrow e,\mu)\approx 6.5\%$, assuming
$L= 2-20$ fb$^{-1}$ as integrated luminosity, in
a window of 10 GeV at $p_T = 100$ GeV, one finds
650--6500 $Z+j$ events through LO, hence a
$\delta\sigma/\sigma\approx -1.5\%$ EW NLO correction corresponds to 10--100 
fewer 
events. At CERN, for the same production and decay channel, assuming now 
$L= 30$ fb$^{-1}$, in a window of 40 GeV at $p_T = 450$ GeV, 
we expect about 1200 $Z+j$ events from LO, so that a 
$\delta\sigma/\sigma\approx -12\%$ EW NLO correction 
corresponds to 140 fewer events. In line with the normalisations seen
in the top frames of  Fig.~\ref{fig:V} and the size of the corrections
in the bottom ones, absolute rates for the photon are similar
to those for the massive gauge boson while ${\cal O}(\alpha_{\rm{S}}\alpha_{\rm{EW}}^2)$ corrections are about a factor of two 
smaller.

Altogether, these results point to the relevance of one-loop
${\cal O}(\alpha_{\rm{S}}\alpha_{\rm{EW}}^2)$ contributions for precision
analyses of prompt-photon and neutral Drell-Yan events at both Tevatron and
LHC, also
recalling that the residual scale dependence of the known
higher order QCD corrections
to processes of the type (\ref{procs_neutral}) is very small 
in comparison \cite{Arnold:1989ub}. Another relevant aspect is
that such higher order weak terms introduce parity-violating
effects in hadronic observables \cite{Ellis:2001ba}, which can be observed 
at (polarised) RHIC-Spin \cite{Bunce:2000uv}.

\end{document}